\documentclass[aps,prb,twocolumn,groupedaddress,amsmath,amssymb,showpacs,10pt,floatfix]{revtex4-1}

\usepackage{amssymb}
\usepackage{amsmath}
\usepackage{graphicx}
\usepackage{epsfig}
\usepackage{longtable}
\usepackage{epstopdf}
\usepackage{dcolumn}
\usepackage{bm}
\usepackage{cancel}
\usepackage[
pdfstartview=XYZ,
bookmarks=false,
colorlinks=true,
linkcolor=blue,
urlcolor=blue,
citecolor=blue,
bookmarks=false,
linktocpage=true, 
hyperindex=false
]{hyperref}
\usepackage{color}

\begin{document}

\title{Electrons and holes in phosphorene}
\author{Pengke Li}
\email{pengke@umd.edu}
\author{Ian Appelbaum} 
\affiliation{Department of Physics and Center for Nanophysics and Advanced Materials, University of Maryland, College Park, MD 20742}

\begin{abstract}
We present a symmetry analysis of electronic bandstructure including spin-orbit interaction close to the insulating gap edge in monolayer black phosphorus (``phosphorene"). Expressions for energy dispersion relation and spin-dependent eigenstates for electrons and holes are found via simplification of a perturbative expansion in wavevector $k$ away from the zone center using elementary group theory. Importantly, we expose the underlying symmetries giving rise to substantial anisotropy in optical absorption, charge and spin transport properties, and reveal the mechanism responsible for valence band distortion and possible lack of a true direct gap. 
\end{abstract}
\maketitle

\section{Introduction}

The experimental isolation of atomically thin 2-dimensional layers from van~der~Waals-bonded 3-dimensional solids has exposed many new opportunities for revealing unconventional electron transport physics. This research area, famously begun with exfoliation of the semimetal graphene from bulk graphite, has now vastly expanded to include work on related group-IV structures (silicene\cite{Aufray_APL2010, *Lalmi_APL2010}, germanene\cite{Davila_arxiv2014}, and stanene\cite{Barfuss_PRL2013}), and on binary semiconductors such as transition-metal dichalcogenides (WS$_2$\cite{Song_PRL2013}, etc.), topological insulators such as Bi$_2$Se$_3$\cite{Hasan_RMP2010}, and the group-III-V insulator boron nitride\cite{Alem_PRB2009}. However, until recently, little has been done to explore the possibility that other forms of elemental compounds beyond group-IV can be exfoliated into few- or single-layer structures like graphite can. Among candidate bulk source materials, orthorhombic black phosphorus (elemental group V) has emerged as a contender.  

The classical literature on this substance, reviewed in  Ref. \onlinecite{Morita_ApplPhysA1986}, is fairly complete at first glance. Single crystals of this ambient-stable allotrope are typically produced using high-pressure Bridgman growth\cite{Bridgman_JACS1914, Maruyama_Physica1981} from which atomic structure was first determined via X-ray diffraction more than 50 years ago.\cite{Hultgren_JChemPhys1935,Brown_ActaCryst1965} Early experimental results on charge transport\cite{Keyes_PR1953} were complemented by investigations into both electrical and optical properties\cite{Warschauer_JAP1963, Narita_Physica1983} and compared with theoretical predictions\cite{Asahina_JPhysC1984, Low_arxiv2014}.  Unexpected phenomena including superconductivity\cite{Kawamura_SSC1984} up to 13~K and evidence of 2D transport\cite{Baba_JPhys1992} motivated a modest resurgence of interest several decades ago. 

Following in the footsteps of graphene's rise in the past decade, more recent experimental work on black phosphorus has focused on field-effect transistor action using thin multi-layered exfoliated flakes as channel conductor.\cite{Koenig_APL2014, Liu_ACSNano2014} \emph{Single layer} black phosphorus, dubbed ``phosphorene", is of particular interest but has not yet been incorporated into electrical devices. The properties of this 2-dimensional semiconductor have been studied in detail only by using the familiar machinery for bandstructure calculation. Several groups have addressed this problem using different approaches, including the empirical tight-binding\cite{Takao_Physica1981,*Takao_JPSJ1981}\cite {Rudenko_PRB2014}, pseudopotential\cite{Asahina_JPSJ1982}, and \textit{ab initio} (Density Functional Theory (DFT))\cite{Du_JAP2010,Xia_NatureComm2014, Qiao_NatureComm2014,Li_NatureNano2014,Rodin_PRL2014} methods, all of which can be compared to empirical dispersion relations obtained using Angle-Resolved Photo-Emission Spectroscopy from the clean surface.\cite{Takahashi_SSC1983} Among the intriguing bandstructure features found are p-type semiconducting bandgap in the visible or infrared region and large excitonic binding energy,\cite{Liu_ACSNano2014,Tran_PRB14} prominent anisotropy of effective mass and hence carrier mobility,\cite{Qiao_NatureComm2014,Xia_NatureComm2014} ultraflat valence band dispersion and possible indirect bandgap, strain-induced gap modification,\cite{Rodin_PRL2014} high optical efficiency, \cite{Buscema_Nano14} etc. 

Despite this abundance of bandstructure results in the available literature, several elementary questions remain unanswered, all of which become crucially important once single-layer phosphorene devices are experimentally realized. For example, what is the origin of the large valence-band effective mass anisotropy? What mechanisms determine whether this material has a truly direct bandgap? What are the optical transition selection rules? What are the dominant wavefunction components that dictate spin-dependent properties? etc. The answers to these questions are essential in providing insight for predictions of the properties of electrons and holes affecting charge and spin transport in this material, and are therefore necessary in developing possible device applications. 

Various brute-force numerical schemes can churn out the relevant quantities needed to answer the questions above, but they often come at the expense of obscuring the physics at their root, i.e. the fundamental symmetries manifest in the structure of this physical materials system. In the present paper, we exploit the discrete lattice/wavefunction symmetries in phosphorene using the formal results of group theory to directly answer these questions. By first identifying the symmetry properties of wavefunctions at the Brillouin zone center, we simplify $\mathbf{k}\cdot\mathbf{\hat p}$ perturbation theory using the method of invariants\cite{BirPikus} and matrix element theorem to identify terms contributing to the dispersion and spin-dependent eigenstates of all relevant bands at nearby momenta. All the symmetry-protected properties are captured by $\mathbf{k}\cdot\mathbf{\hat p}$ parameters that can be easily verified by numerical calculations and empirically determined by further experiments.

The present paper is organized as follows: Section~\ref{sec:general} provides general information on the symmetry of the phosphorene 2D lattice, from which we analyze the symmetries of constituent atomic orbitals and the nearly free electron model. In Section~\ref{sec:III}, we investigate the spin-independent part of the Hamiltonian using the method of invariants and reveal the fundamental origins of the effective mass anisotropy, especially those interactions resulting in an ultraflat valence band. Optical selection rules are also provided. Using the same approach, in Section~\ref{sec:IV} we focus on the spin-orbit interaction and construct the spin-dependent eigenstates. From these results, we analyze spin relaxation anisotropy before providing a summary and outlook in  Section~\ref{sec:V}.

\section{Symmetry Considerations}\label{sec:general}

\subsection{Lattice symmetry and space group operators}

\begin{figure}
\includegraphics[width=8.cm] {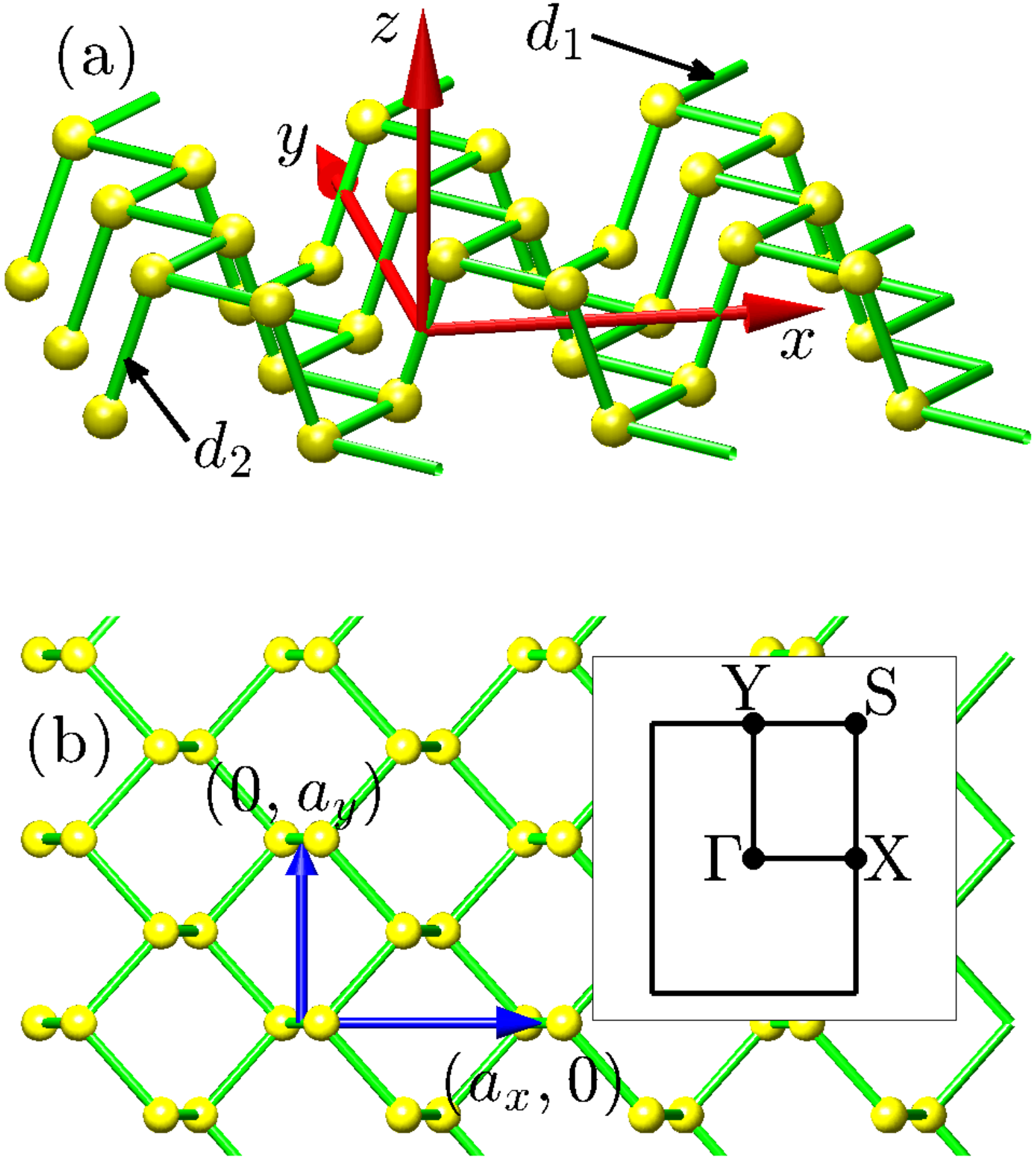}
\caption{ Orthorhombic lattice of phosphorene. (a) In real-space, two types of bonds between neighboring atoms are indicated by $d_1$ and $d_2$. The origin of the Cartesian coordinates (red arrows) is chosen to be at the center of a $d_2$ bond. (b) Top view, showing the two lattice vectors (blue arrows). Inset: 2D projection of the reciprocal lattice Brillouin zone, with high symmetry points indicated.
\label{fig:lattice}}
\end{figure}

The black phosphorus monolayer is a 2D hexagonal lattice that is buckled, or ``puckered", along the armchair direction. This geometry results in two types of bonds [see Fig.~\ref{fig:lattice}(a)]. P atoms connected by bonds parallel to the 2D plane (Type I, with bond length $d_1=2.224$ \AA) form upper and lower sublayers, while bonds connecting P atoms between these two sublayers (Type II) have a length $d_2=2.244$ \AA~and are oriented $71.7^{\circ}$ out of the plane. In the following discussion, we use Cartesian coordinates with the origin at the center of one Type II bond. The $z$ axis is chosen to be out of plane, with the in-plane $x$ axis along the armchair direction and the $y$ axis transverse to it. In this rectangular $x-y$ basis, the two Bravais lattice constants are $a_x = 4.376$\AA~and $a_y = 3.314$\AA  ~[see Fig.~\ref{fig:lattice}(b)], and within a unit cell there are four P atoms. The inset in Fig.~\ref{fig:lattice}(b) shows the first Brillouin zone of this 2D orthorhombic lattice with high symmetry points labeled: $\Gamma$-point is the zone center, and $\mathrm{X}$ and $\mathrm{Y}$ points are at $(\frac{\pi}{a_x},0)$ and $(0,\frac{\pi}{a_y})$, respectively, half of the reciprocal lattice vectors.

Phosphorene shares the same in-plane translation symmetry with its bulk counterpart black phosphorus, whose space group is base-centered orthorhombic with international number and symbol $64:Cmca$.\cite{BradleyCracknell} The lattice structure in Fig.~\ref{fig:lattice} provides all necessary information about its nonsymmorphic space group, whose factor group is isomorphic to the point group $D_{2h}$. There are eight elements in this nontrivial factor group; each of them is a coset about the direct product of a symmetry operator and the lattice vector translation subgroup $T$.  Among these eight symmetry operators, four are pure rotations (either proper or improper), including the identity operator $E$, the space inversion operator $i$, the operator corresponding to $180^{\circ}$ rotation around the $y$ axis $C_{2y}$, and the  operator corresponding to reflection with respect to the $y=0$ plane $R_y$. The remaining four are a translation $\tau =(\frac{a_x}{2},\frac{a_y}{2})$ in addition to pure rotations, including $\tau C_{2x}$, $\tau C_{2z}$, $\tau R_x$ and $\tau R_z$. Since the $D_{2h}$ group is abelian (commutative), each group element forms a single class.

The most important information about the symmetry of this space group and its eight irreducible representations (IRs) is included in the character table (see Table~\ref{tab:CharacterTable}). Since this is the same group as that of the Brillouin zone center $\Gamma$-point, these IRs are denoted by $\Gamma_i^{+(-)}$, with subscript $i = 1,2,3, 4$, and superscript $+$ ($-$) indicating even (odd) parity under the inversion operator. In this table we also list the basis functions of some IRs. Here, $x$, $y$ and $z$ are components of a polar vector (e.~g. the momentum operator $\mathbf{\hat p}$) while $A_x$, $A_y$ and $A_z$ are those of an axial, or ``pseudo-", vector (essential for analyzing the effect of spin-orbit interaction with vector field $\propto \nabla V \times \mathbf{\hat p}$). 

\begin{table}
\caption{Character table of the $\Gamma$-point space group, including basis functions of the IRs. ${x}$, ${y}$, and ${z}$ are components of polar vectors, while ${A}_x$, ${A}_y$ and ${A}_z$ are components of axial vectors.}
\label{tab:CharacterTable}
\renewcommand{\arraystretch}{1}
\begin{tabular}{c|cccc|cccc|c}
\hline \hline
&$E$&$\tau C_{2x}$&$C_{2y}$&$\tau C_{2z}$&$i$&$\tau R_x$ &$R_y$&$\tau R_z$&\\ \hline
$\Gamma_1^+$  & 1 &1& 1& 1 &1 &1 &1 &1&\\
$\Gamma_2^+$  & 1 &-1&1 &-1 &1 &-1&1 &-1 & ${A}_y$\\
$\Gamma_3^+$  & 1 &1&-1 &-1 &1 &1&-1 &-1&${A}_x$\\
$\Gamma_4^+$  & 1 &-1& -1& 1 &1 &-1 &-1 &1&${A}_z$\\\hline
$\Gamma_1^-$  & 1 &1&1 &1 &-1 &-1&-1 &-1&\\
$\Gamma_2^-$  & 1 &-1&1 &-1 &-1 &1&-1 &1&${y}$\\
$\Gamma_3^-$  & 1 &1&-1 &-1 &-1 &-1&1 &1&${x}$\\
$\Gamma_4^-$  & 1 &-1&-1 &1 &-1 &1&1 &-1&${z}$\\
\hline
\end{tabular}
\end{table}

In the following subsection, we will briefly discuss the symmetries of the atomic orbitals and the empty lattice ``nearly free" electron band structure at the $\Gamma$-point, using the information in Table~\ref{tab:CharacterTable}. These considerations provide intuition on the symmetry-related coupling of the eigenstates near the zone center and are important for understanding the overall electronic structure.

\subsection{Atomic orbital symmetries at the center of the Brillouin zone}
\label{sec:IIB}

The band structure of black phosphorus was studied by Takao \textit{et. al.} \cite{Takao_Physica1981,*Takao_JPSJ1981} with a spin-independent tight-binding model using a basis of one $s$ orbital and three $p$ orbitals. With four P atoms within a unit cell, there are therefore sixteen bands. By calculating the hopping energy and orbital overlaps between neighboring atoms, they determined the band gaps of black phosphorus in the form of monolayer and bulk, and indexed the symmetries of eigenstates at the $\Gamma$-point. In general, wavefunctions constructed this way will consist of $sp^3$ hybridized atomic orbitals. However, at the zone center, the $p_y$ orbital (which is odd under $R_y$) is isolated from the remaining $s$, $p_x$ and $p_z$ orbitals that remain mixed (similar to the case of pure $p_z$ orbitals in monolayer graphene due to the $R_z$ operator). 

\begin{figure}
\includegraphics[width=8cm] {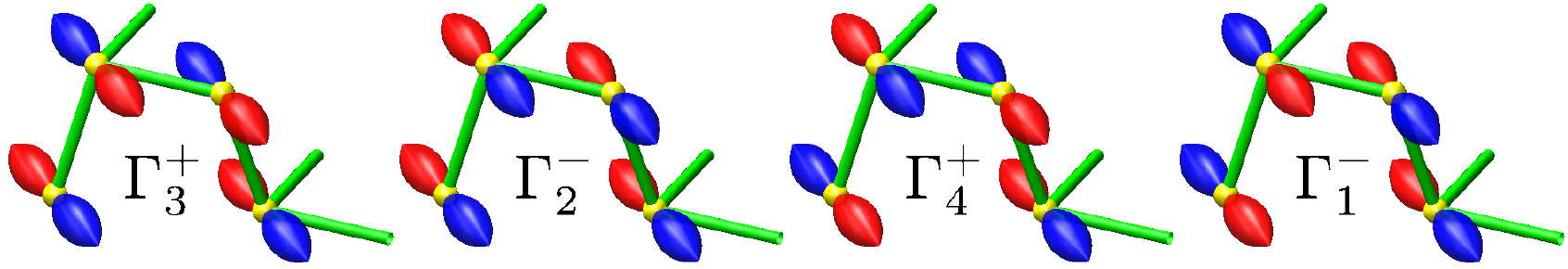}
\caption{Four configurations of the $p_y$ orbitals within a unit cell, together with their associated IRs, all of which are odd under the reflection operator $R_y$. They represent the four zone-center eigenstates composed solely from the pure $p_y$ atomic orbitals. From left to right, the eigenenergies of the four orbital states decrease [$E(\Gamma_3^+)>E(\Gamma_2^-)>E(\Gamma_4^+)>E(\Gamma_1^-)$], according to the bonding or antibonding nature of the covalent bonds between neighboring atoms. Note that the $\Gamma_2^-$ configuration also represents the long-wavelength acoustic phonon mode where all atoms move in-phase along the zigzag direction.
\label{fig:py}}
\end{figure}

According to the character table (Table~\ref{tab:CharacterTable}), all the sixteen eigenstates at the $\Gamma$-point are nondegenerate, since all the IRs are one-dimensional. Four of these IRs, associated with the four $p_y$ orbital configurations illustrated in Fig.~\ref{fig:py}, are odd under the reflection operator $R_y$: $\Gamma_3^+$, $\Gamma_4^+$, $\Gamma_1^-$ and $\Gamma_2^-$. The bonding energy is dominated by Type I bonds (within each sublayer) that hybridize $pp\pi$ and $pp\sigma$ covalent chemical bonds, the latter of which is much stronger due to higher orbital overlap. Type II bonds (connecting the two sublayers) are pure but weaker $pp\pi$ bonds and have only a secondary contribution to the bonding energy. By considering the bonding and antibonding nature of the $p_y$ orbitals, we give in Fig.~\ref{fig:py} the relative order of the four eigenenergies, which matches the tight-binding calculation \cite{Takao_Physica1981,*Takao_JPSJ1981}.

\begin{figure}
\includegraphics[width=8cm] {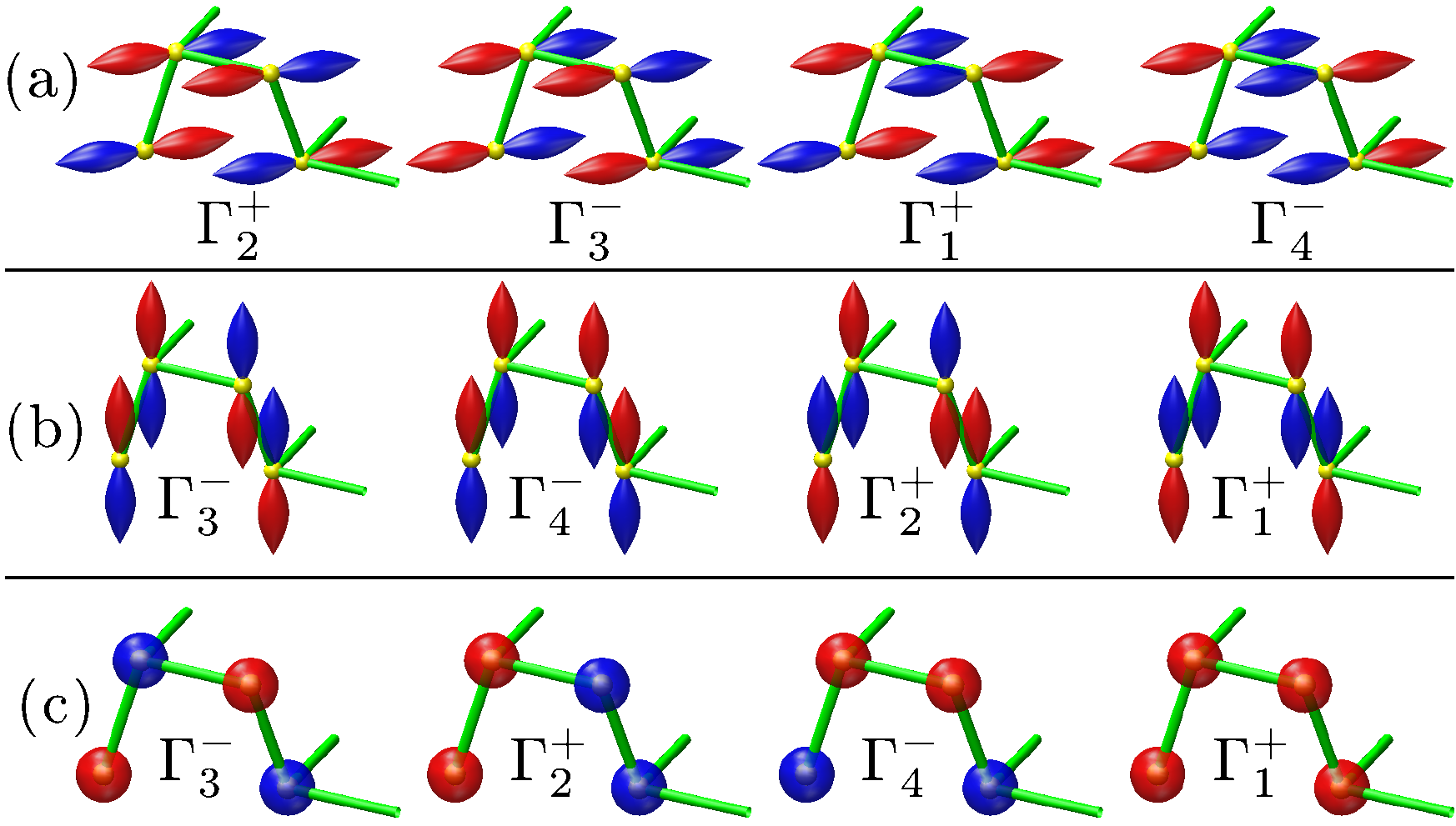}
\caption{ Same as Fig.~\ref{fig:py}, here for (a) $p_x$, (b) $p_z$ and (c) $s$ orbitals that are even under the reflection operator $R_y$. IRs in each row are listed from left to right in descending order of total covalent bond energy. Similar to $\Gamma_2^-$ in Fig.~\ref{fig:py}, here we see configurations corresponding to the remaining two long wavelength acoustic phonon modes: $\Gamma_3^-$ in (a) corresponds to in-plane motion along the armchair direction while $\Gamma_4^-$ in (b) corresponds to out-of-plane (flexural) phonons.
\label{fig:pxpzs}}
\end{figure}

Takao \textit{et. al.} gave the $p_z$ orbital configurations at the bandgap edge ($\Gamma_2^+$ for the valence band and $\Gamma_4^-$ for the conduction band). In Fig.~\ref{fig:pxpzs}(a), (b), and (c), we list the configurations of all the $p_x$, $p_z$ and $s$ orbitals, respectively, for the four IRs $\Gamma_1^+$, $\Gamma_2^+$, $\Gamma_3^-$ and $\Gamma_4^-$ that are even under the reflection operator $R_y$, and order them according to their bond energy. Within the $sp^3$ tight-binding model, each of these four representations corresponds to three bands, among which the contributions of $p_x$, $p_z$ and $s$ orbital components with the same symmetry vary. 

Similar to our discussion of the $p_y$ orbital, we now examine the bonding or antibonding characteristics of these atomic orbital configurations and determine the relative energy ordering of the eigenstates. In the following sections we will focus on the band edge states which belong to $\Gamma_2^+$ and $\Gamma_4^-$, both dominated by the $p_z$ orbital. In this case [see Fig.~\ref{fig:pxpzs}(b)], $pp\sigma$ bonds are within the Type II bonds, which are bonding (antibonding) in $\Gamma_2^+$ ($\Gamma_4^-$) with lower (higher) energy assigning it the top of the valence band (bottom of the conduction band).

Using the first-principles calculation package {\sc Quantum ESPRESSO}\cite{QE-2009}, we studied this energy order of the $\Gamma$-point eigenstates. Different density functionals and pseudopotentials chosen as input to the \textit{ab initio} calculation vary the detailed values of energy differences at the $\Gamma$ point as well as the dispersion curves away from it. However, the band ordering at the $\Gamma$ point is consistent with those given by Takao \textit{et. al.} (only a few bands very close in energy switch places). This verifies that the order of the $\Gamma$ point eigenstates is generally determined by the symmetries of the atomic orbitals alone.

\subsection{Empty-lattice band structure \label{sec:IIC}}
The atomic orbitals give a perturbative picture of the electronic structure in the tight-binding regime. A useful approximation in the opposite (delocalized) extreme is the nearly-free electron model that describes the empty-lattice band structure, in which the electronic states are pure planewaves with wave numbers given by the reciprocal lattice vectors ($\mathbf{G_n}$). Using projection operators, we obtain the symmetrized wavefunctions which are linear combinations of planewaves degenerate at the $\Gamma$-point. When the atomic potentials are introduced into this nearly free electron model, only those symmetrized wavefunctions belonging to the same IR can be linearly combined to form the real eigenstates (the symmetry of the eigenstates from empty-lattice to the real lattice is maintained) and result in broken degeneracy with the possibility of bandgaps. 

An important potential concern with this approach is that it does not account for the transformation properties of the wavefunctions under out-of-plane reflection. Therefore, in principle, wavefunctions constructed in this way provide only incomplete symmetry properties. However, since the dynamical characteristics of interest involve only in-plane symmetries, the simple nearly-free electron model can indeed provide sufficient information. 

\begin{figure}
\includegraphics[width=8cm] {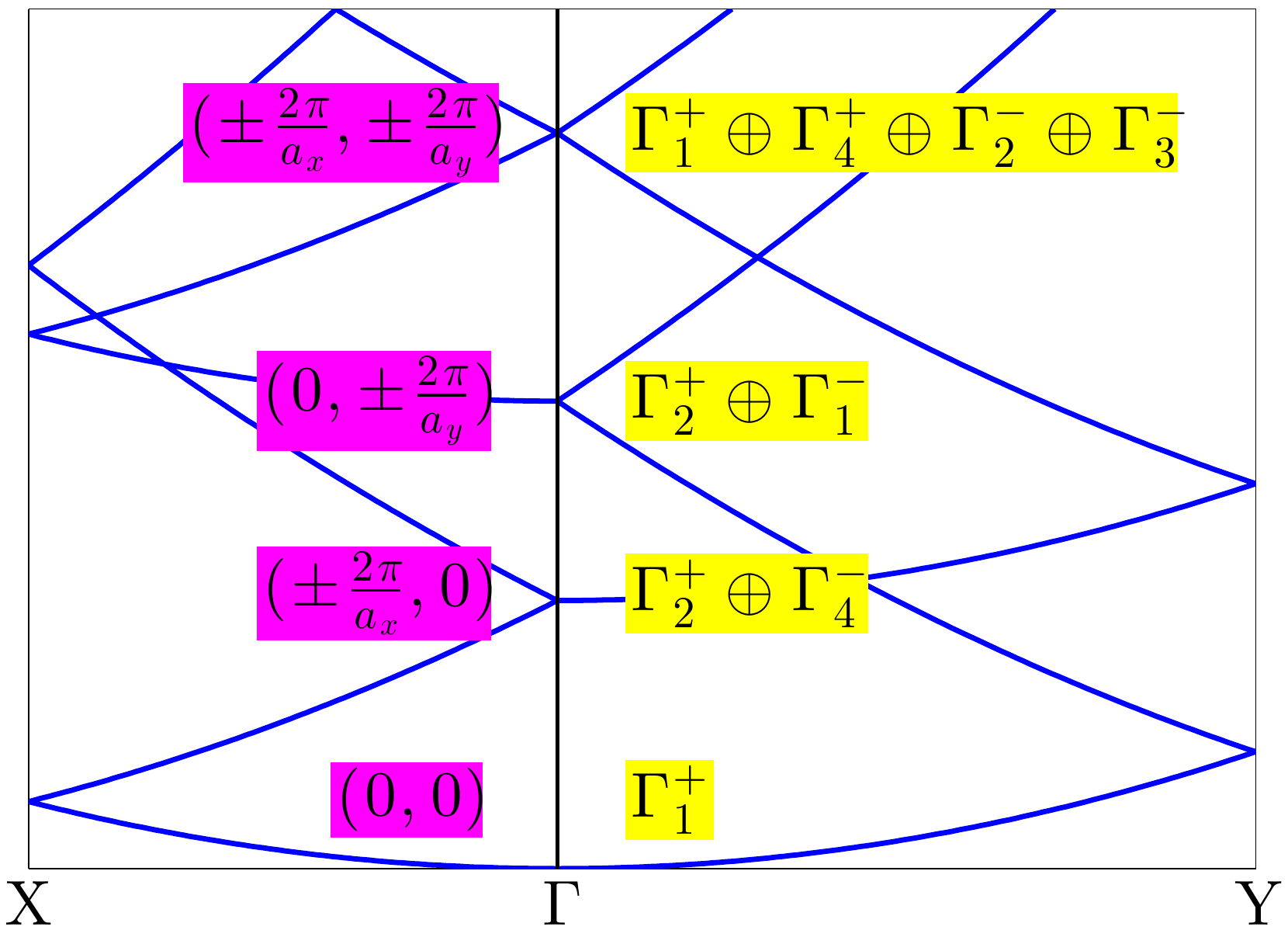}
\caption{Free-electron bandstructure, including symmetry labels (yellow) of zone-center eigenstates formed from superpositions of plane-waves centered at the reciprocal lattice points indicated (purple).
\label{fig:empty_lattice}}
\end{figure}

Because of the orthorhombic symmetry of monolayer black phosphorus, the degeneracy of nearly-free electron states at the $\Gamma$-point can only be singlet ($\mathbf{G_n}=0$), doublet ($\mathbf{G_n}$ on the $k_x$ or $k_y$ axes) or quartet (general $\mathbf{G_n}$). Fig.~\ref{fig:empty_lattice} shows the empty-lattice band structure including the lowest nine $\Gamma$-point eigenstates. In the following sections, we will show that the in-plane momentum matrix elements $\langle\hat p_{x,y}\rangle$ of the zone center eigenstates play a fundamental role  in determining the electronic structure of phosphorene. The nearly-free electron model explicitly shows that $\langle \hat p_{x,y}\rangle$ are nonzero only between degenerate states with relevant wavevector components. For instance, for the doubly degenerate first excited states in the form of symmetrized wavefunctions with $\mathbf{G_n} = (\pm\frac{2\pi x}{a_x},0)$, we have
\begin{equation}
\langle\Gamma_2^+|\hat p_x |\Gamma_4^-\rangle= \langle\cos\frac{2\pi x}{a_x}|\frac{\hbar}{i}\frac{\partial}{\partial x} |\sin\frac{2\pi x}{a_x}\rangle = \frac{2\pi \hbar}{a_x},
\label{eq:px_PW}
\end{equation}
while $\langle\Gamma_2^+|\hat p_y |\Gamma_4^-\rangle=0$. Similarly, $\hat p_x$ and $\hat p_y$ between either of these two states and any other remaining $\Gamma$-point planewave state vanish, consisent with the matrix element theorem as discussed below.

\section{Spin-independent properties}\label{sec:III}

In this section, we study the spin-independent band structure properties close to the bandgap in phosphorene. We construct the $\mathbf{k}\cdot\mathbf{\hat p}$ Hamiltonian near the zone center using the method of invariants,\cite{BirPikus} and show that the dispersion relation of the conduction band and valence band can be captured by an effective mass approximation only in the $k_x$ direction, while in the $k_y$ direction the unique coupling from remote bands can potentially lead to an indirect bandgap.

\subsection{Hamiltonian and method of invariants}

Following conventional $\mathbf{k}\cdot\mathbf{\hat p}$ theory, the Hamiltonian is given by
\begin{align}
H= H_0+H_{\mathbf{k}\cdot\mathbf{\hat p}}+H_{\text{SO}}+H_{\text{SO},{\mathbf{k}}},
\label{eq:total_H}
\end{align}
where 
\begin{align}
&H_0 = \frac{\hbar^2}{2m_0}(k_x^2+k_y^2),\label{eq:H_0}\\
&H_{\mathbf{k}\cdot\mathbf{\hat p}} = \frac{\hbar}{m_0}(k_x \hat p_x+k_y \hat p_y),\label{eq:H_kp}\\
&H_{\text{SO}} = \frac{\hbar}{4m_0^2c^2}\nabla V\times\mathbf{\hat p}\cdot\vec{\sigma},{\text{ and}}\label{eq:H_SO}\\
&H_{\text{SO},{\mathbf{k}}} = \frac{\hbar^2}{4m_0^2c^2}\left[(k_x\sigma_y-k_y\sigma_x)\frac{\partial V}{\partial z}+\right.\nonumber\\
& \qquad \qquad \qquad \quad \left.k_y\sigma_z\frac{\partial V}{\partial x}-k_x\sigma_z\frac{\partial V}{\partial y}\right].  \label{eq:H_SOk} 
\end{align}
Here, $H_0$ is the in-plane free electron dispersion, and $H_{\mathbf{k}\cdot\mathbf{\hat p}}$ is the $\mathbf{k}\cdot\mathbf{\hat p}$ term to be treated perturbatively. In this section we will focus on these two spin-independent terms. The spin-related properties are captured by the $\mathbf{k}$-independent $H_{\text{SO}}$ and $\mathbf{k}$-dependent $H_{\text{SO},{\mathbf{k}}}$ terms, and are discussed in the subsequent section. Note that for light atoms like phosphorus, the interaction strength hierarchy is generally $H_{\mathbf{k}\cdot\mathbf{\hat p}}\gg H_{\text{SO}} \gg H_{\text{SO},{\mathbf{k}}} $.

\begin{table} 
\caption{Table of invariants}
\label{tab:table_of_inv}
\renewcommand{\arraystretch}{1.2}
\begin{tabular}{c|c|c|c|c|c|c|c}
\hline \hline
IRs & $\Gamma_1^+$ &$\Gamma_2^+$ & $\Gamma_3^+$ & $\Gamma_4^+$& $\Gamma_2^-$&$\Gamma_3^-$ &$\Gamma_4^-$\\\hline
Invariants &$k_x^2+k_y^2$ &  $\sigma_y$   &  $\sigma_x$   &  $\sigma_z$  &  $
\begin{array}{c}
k_y,\\
-k_x\sigma_z
\end{array}
$ &  $
\begin{array}{c}
k_x,\\
k_y\sigma_z
\end{array}
$ & $k_x\sigma_y-k_y\sigma_x$ \\\hline
\end{tabular}
\end{table}

Given the symmetry of the problem, we naturally choose to represent this Hamiltonian in a basis defined by the spin-independent $\Gamma$-point eigenstates. Matrix elements of perturbative terms in Eq.~(\ref{eq:total_H}) can then be determined by the method of invariants\cite{BirPikus}: only if the IR associated with the invariant component operator is included in the direct sum decomposition of the direct product of the two IRs of the basis functions can the matrix element be nonzero. In Table~\ref{tab:table_of_inv}, we list the association of the IRs with all the invariant components of terms in Eq.~(\ref{eq:total_H}) according to their transformation properties under the symmetry operators in Table~\ref{tab:CharacterTable}.

Our main focus is on the lowest conduction band and the highest valence band, which belong to the IRs $\Gamma_4^-$ and $\Gamma_2^+$, respectively. For convenience, we list the direct product of these two IRs with all IRs in Table ~\ref{tab:directproduct}. In particular, we will reveal the origin of the anisotropy of the energy dispersion relation in the $k_x$ and $k_y$ directions, which is naturally endowed by the orthorhombic symmetry of the crystal lattice.

\begin{table}
\caption{Direct product between $\Gamma_2^+$ ($\Gamma_4^-$) and all IRs.}
\label{tab:directproduct}
\renewcommand{\arraystretch}{1.2}
\begin{tabular}{c|cccccccc}
\hline \hline
 &$\Gamma_1^+$  &$\Gamma_2^+$  &$\Gamma_3^+$  &$\Gamma_4^+$  &$\Gamma_1^-$  &$\Gamma_2^-$  &$\Gamma_3^-$  &$\Gamma_4^-$  \\ \hline
$\Gamma_2^+$  & $\Gamma_2^+$  &$\Gamma_1^+$&$\Gamma_4^+$ &$\Gamma_3^+$ &$\Gamma_2^-$ &$\Gamma_1^-$&$\Gamma_4^-$ &$\Gamma_3^-$ \\\hline
$\Gamma_4^-$  & $\Gamma_4^-$  &$\Gamma_3^-$&$\Gamma_2^-$ &$\Gamma_1^-$ &$\Gamma_4^+$ &$\Gamma_3^+$&$\Gamma_2^+$ &$\Gamma_1^+$\\
\hline
\end{tabular}
\end{table}

\subsection{Effective mass of electrons and holes in the \texorpdfstring{$k_x$}{kx} direction}

\begin{figure}
\includegraphics[width=8.cm] {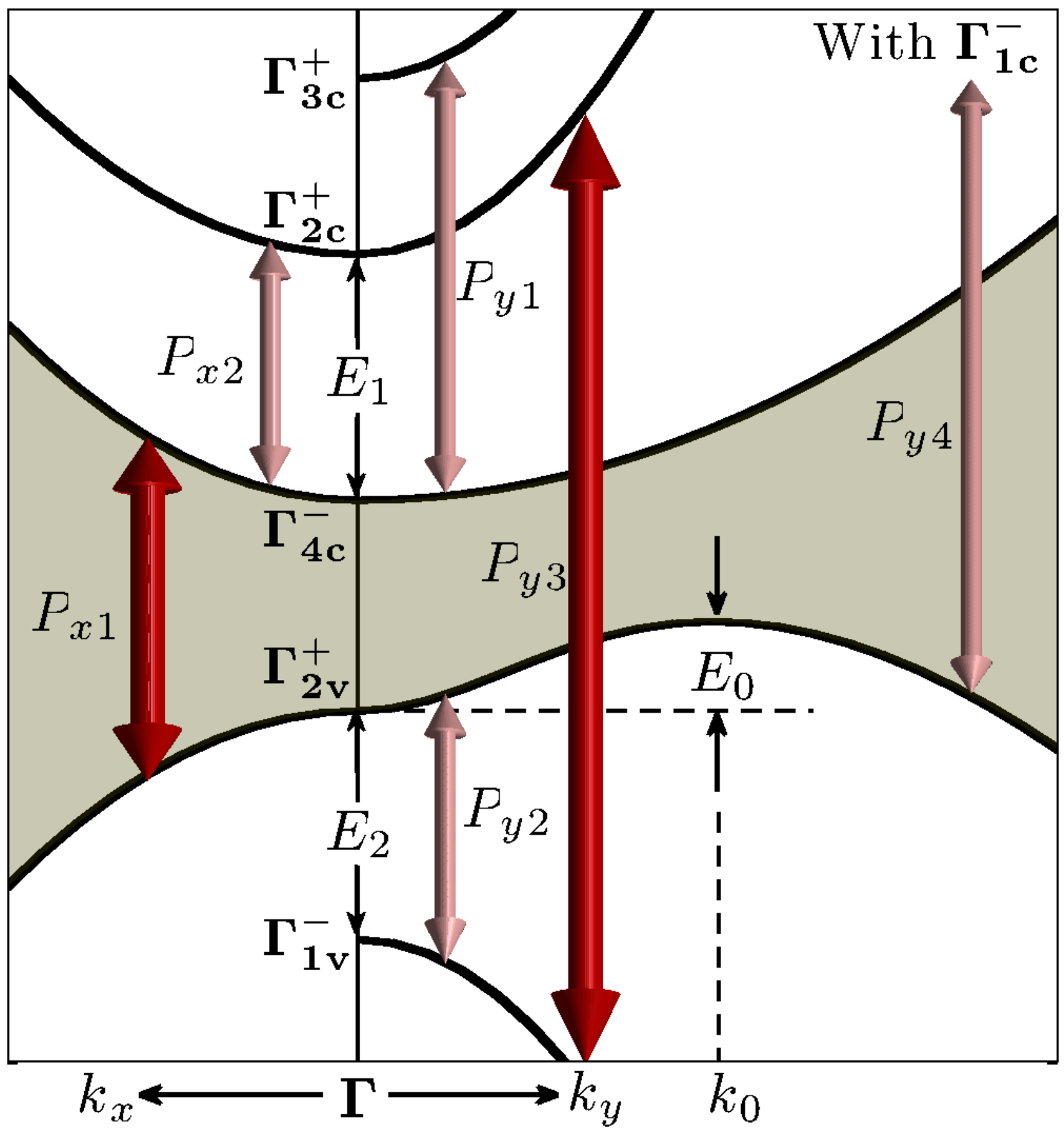}
\caption{Schematic bandstructure of phosphorene near the zone center, illustrating the relevant perturbative interactions allowed by symmetry. Red double arrows indicate dominant coupling terms, and pink arrows highlight important non-negligible additional interactions. The gray region represents the forbidden gap. See Eq. (\ref{eq:E_h_ky}) and related text for conditions resulting in indirect gap ($E_0, k_0\neq 0$). \emph{The plotting of bands are not to scale, especially $E_0$ is exaggerated for better illustration. }
\label{fig:bands}}
\end{figure}

With the help of Table~\ref{tab:table_of_inv}, we see that the perturbative term proportional to $k_x\hat p_x$ in  Eq.~(\ref{eq:H_kp}) belongs to $\Gamma_3^-$. This term directly couples the lowest conduction band and the highest valence band in off-diagonal matrix elements ($\Gamma_3^- = \Gamma_4^-\otimes\Gamma_2^+$; see Table~\ref{tab:directproduct}).  Similarly, the term proportional to $k_x^2$ in Eq.~(\ref{eq:H_0}) belongs to $\Gamma_1^+$ and corresponds to diagonal matrix elements $\Gamma_4^-\otimes\Gamma_4^-$ and $\Gamma_2^+\otimes\Gamma_2^+$. Notice that the second lowest conduction band also belongs to $\Gamma_2^+$ and therefore is expected to affect the $\Gamma_4^-$ bottom conduction band significantly as well. Given the fact that all other remote bands belonging to $\Gamma_2^+$ or $\Gamma_4^-$ are much further away in energy, we can describe the $k_x$ direction dispersion relation close to the bandgap by a minimal $3\times3$ Hamiltonian with a basis of $\{\Gamma_{2\text{c}}^+,\Gamma_{4\text{c}}^-,\Gamma_{2\text{v}}^+\}$ (where the subscripts $\text{v}$ and $\text{c}$ indicate `valence' and `conduction', respectively):

\begin{eqnarray}
H_{3\times3} &=& \left(
\begin{array}{ccc}
E_1+E_g+\frac{\hbar^2k_x^2}{2m_0}   &   P_{x2} k_x    &   0   \\
P_{x2} k_x               & E_g+\frac{\hbar^2k_x^2}{2m_0} &P_{x1} k_x  \\
0 & P_{x1} k_x & \frac{\hbar^2k_x^2}{2m_0}
\end{array}
\right),
\label{eq:H_3by3}
\end{eqnarray}
where $E_g$ is the bandgap at the $\Gamma$-point, $E_1$ is the energy difference between $\Gamma_{2\text{c}}^+$, and $\Gamma_{4\text{c}}^-$, and the off-diagonal matrix elements are dependent on
\begin{align}
&P_{x1} = \frac{\hbar}{m_0}\langle \Gamma_{4\text{c}}^-| \hat{p}_x |\Gamma_{2\text{v}}^+\rangle, {\text{ and}} \\
&P_{x2} = \frac{\hbar}{m_0}\langle \Gamma_{4\text{c}}^-| \hat{p}_x |\Gamma_{2\text{c}}^+\rangle.
\label{eq:Px}
\end{align}
Such couplings are shown by double arrows in Fig.~\ref{fig:bands} on the $k_x$ side of the $\Gamma$-point in a schematic bandstructure. Note that, for zone center wavefunctions, we can always adjust their overall phases so that $P_{x1}$ and $P_{x2}$ are real numbers.

Applying the L\"owdin partitioning method\cite{Lowdin_JCP1951} to lowest order, Eq.~(\ref{eq:H_3by3}) yields analytic expressions for the dispersion relations of electrons and holes in the $k_x$ direction:

\begin{align}
E_e(k_x) &= E_g+\frac{\hbar^2}{2m_0}k_x^2-\frac{P_{x2}^2}{E_{1}} k_x^2+\frac{P_{x1}^2}{E_{g}} k_x^2, {\text{ and}}
\label{eq:E_e_kx} \\
E_h(k_x) &= \frac{\hbar^2}{2m_0}k_x^2-\frac{P_{x1}^2}{E_{g}}k_x^2.
\label{eq:E_h_kx} 
\end{align}
From these eigenenergies, we can then calculate the effective masses via $\left[\frac{1}{\hbar^2}\frac{d^2E}{dk_x^2}\right]^{-1}$:
\begin{align}
\frac{1}{m_{e,x}} &= \frac{1}{m_0}-\frac{2P_{x2}^2}{E_{1}}+\frac{2P_{x1}^2}{E_{g}},{\text{ and}}
\label{eq:m_e_kx} \\
\frac{1}{m_{h,x}} &= -\frac{1}{m_0}+\frac{2P_{x1}^2}{E_{g}}.
\label{eq:m_h_kx} 
\end{align}

To evaluate $m_e$ and $m_h$, we utilize the results of the free electron model in Sec.~\ref{sec:IIC}. Our DFT calculations show that both $\Gamma_{4\text{c}}^-$ and $\Gamma_{2\text{v}}^+$ have a dominant first excited state planewave component with $\mathbf{G}=(\frac{2\pi}{a_x},0)$, while for $\Gamma_{2,\text{c}}^+$ this component is small and the majority is the second excited state with $\mathbf{G}=(0,\frac{2\pi}{a_y})$, regardless of the DFT details. Thus, the magnitudes of both $m_{e,x}$ and $m_{h,x}$ are largely dictated by the term related to $P_{x1}$ [illustrated in Fig.~(\ref{fig:bands}) by the large red double arrow on the $k_x$ side], with amplitude on the order of $\frac{2\pi\hbar^2}{m_0 a_x}$ [see Eq.~(\ref{eq:px_PW})], while the $P_{x2}$-related term is a small correction to $m_{e,x}$ that reduces the difference between $m_{e,x}^{-1}$ and $m_{h,x}^{-1}$.

It should be noted that a recently-proposed two-band model yields a similar dispersion relation in the $k_x$ direction.\cite{Rodin_PRL2014} In that model, it is claimed that the effect of the  $\Gamma_{2\text{c}}^+$ conduction band (as well as the contribution of remote bands) has been lumped into off-diagonal matrix elements by terms quadratic in $k$ after L\"owdin partitioning. We are compelled to point out that this result is flawed, however, since $k_x^2$ is not an invariant of $\Gamma_4^-\otimes\Gamma_2^+=\Gamma_3^-$, and L\"{o}wdin partitioning always maintains the underlying symmetry. Furthermore, inclusion of specious odd-power terms in the eigenenergies (due to the coupling of the off-diagonal linear and quadratic terms) breaks time-reversal symmetry such that dispersions in $+k_x$ and $-k_x$ directions are unphysically different. For the same reason, the off-diagonal $k_y^2$ term in that two-band model (leading to a $k_y^4$ dispersion relation) is also incorrect. In the following subsection, we will show the correct origins of the flat valence band and possible indirect bandgap in the $k_y$ direction.

\subsection{Dispersion relation in the \texorpdfstring{$k_y$}{ky} direction}

Unlike the case in the $k_x$ direction, direct coupling of the band edge states $\Gamma_{2\text{v}}^+$ and $\Gamma_{4\text{c}}^-$ is absent in the $k_y$ direction; energetically-remote bands must therefore be taken into account. The situation for the conduction band is relatively simple: according to Tables~\ref{tab:table_of_inv}~and~\ref{tab:directproduct}, the invariant component $k_y$ belongs to $\Gamma_2^-$ and couples $\Gamma_{4\text{c}}^-$ to $\Gamma_{3}^+$ states. As discussed in Sec.~\ref{sec:IIB}, within the $sp^3$ tight-binding model giving the lowest sixteen bands, $\Gamma_{3}^+$ is the highest energy state among the four pure $p_y$ orbitals. Numerical calculation shows that it lies beyond $\Gamma_{4\text{c}}^-$  \cite{Takao_Physica1981,*Takao_JPSJ1981}, consistent with our DFT calculation. Together with other (even higher) states with the same symmetry, $\Gamma_{3\text{c}}^+$ repels the dispersion of the lowest conduction band downward via the matrix element
\begin{align}
P_{y1} = \frac{\hbar}{m_0}\langle \Gamma_{3\text{c}}^+| \hat{p}_y |\Gamma_{4\text{c}}^-\rangle.
\label{eq:Py1}
\end{align}
This interaction is schematically shown with a double-sided pink arrow in Fig.~\ref{fig:bands}. Since the planewave component of $\mathbf{G}=(0,\frac{2\pi}{a_y})$ in $\Gamma_{4\text{c}}^+$ is relatively small, $P_{y1}$ is not large enough to reverse the conduction band's positive curvature. However, it does result in a value of the effective mass larger than $m_0$.

The valence band state $\Gamma_{2\text{v}}^+$ requires a different analysis. According to Table~\ref{tab:directproduct}, and reflected in Fig.~\ref{fig:bands}, $\Gamma_{1\text{v}}^-$ can  directly couple to $\Gamma_{2\text{v}}^+$ via $k_y\hat{p_y}$ perturbation. This state, the lowest among the four pure $p_y$ orbital states (see Fig.~\ref{fig:py}), compels us to consider a matrix element
\begin{align}
P_{y2} = \frac{\hbar}{m_0}\langle \Gamma_{1\text{v}}^-| \hat{p}_y |\Gamma_{2\text{v}}^+\rangle.
\label{eq:Py2}
\end{align}
Compared with $P_{x1}$, this quantity is relatively small due to the minority $(0,\frac{2\pi}{a_y})$ planewave component in $\Gamma_{2\text{v}}^+$ (despite its dominant role in $\Gamma_{1\text{v}}^-$). However, without a counterbalancing interaction, its presence would repel the $\Gamma_{2,\text{v}}^+$ band upward, leading to an electron-like positive effective mass and close the bandgap. 

Previous DFT calculation by others has already shown that monolayer black phosphorus potentially posseses an indirect bandgap, in which the maximum of the valence band is located along the $k_y$ direction away from the zone center.\cite{Rodin_PRL2014} Using various input density functionals and pseudopotentials, we have verified the persistence of this feature which leads to a small positive energy difference ($E_0\sim$ meV, depending on the details of the numerical procedures) between the valence band maximum at wavevector $k_0$ ($\approx$ 10\% from BZ edge) and eigenenergy at the zone center. In contrast to an ordinary quadratic dispersion relation, understanding such unusual band structure requires careful examination of the couplings between $\Gamma_{2\text{v}}^+$, $\Gamma_{1\text{v}}^-$, $\Gamma_{2\text{c}}^+$, and further upper conduction bands of $\Gamma_1^-$ symmetry.

We have mentioned that for both $\Gamma_{1\text{v}}^-$ and $\Gamma_{2\text{c}}^+$ states, planewave component $(0,\frac{2\pi}{a_y})$ dominates the wavefunctions. As indicated in Fig.~\ref{fig:bands} by a large red double arrow, the direct coupling of these two states by $k_y\hat{p_y}$ in
\begin{align}
P_{y3} = \frac{\hbar}{m_0}\langle \Gamma_{1\text{v}}^-| \hat{p}_y |\Gamma_{2\text{c}}^+\rangle\sim\frac{2\pi\hbar^2}{m_0 a_y}
\label{eq:Py3}
\end{align}
is very large. Thus, the $\Gamma_{1\text{v}}^-$ band is strongly repelled downward along the $k_y$ direction, giving a dispersion 
\begin{align}
E_{1v}(k_y) &\approx -E_2-\frac{P_{y3}^2}{E_1+E_{g}+E_2}k_y^2,
\label{eq:E_1v} 
\end{align}
where $E_2$ is the energy difference from $\Gamma_{2\text{v}}^+$ to $\Gamma_{1\text{v}}^-$. Both the free electron dispersion and  the effect from $\Gamma_{2\text{v}}^+$ [Eq.~(\ref{eq:Py2})] are neglected in this expression, due to their minor contributions compared with the influence of $P_{y3}$.
As $k_y$ increases, the energy difference between $\Gamma_{2\text{v}}^+$ and $\Gamma_{1\text{v}}^-$ therefore quickly grows, further diminishing their coupling. 

An additional interaction is still needed to induce a hole-like negative effective mass for $\Gamma_{2\text{v}}^+$ to preserve the open bandgap. This role is played by conduction band states with $\Gamma_1^-$ symmetry, as shown in Fig.~\ref{fig:bands} by the matrix element
\begin{align}
P_{y4} = \frac{\hbar}{m_0}\langle \Gamma_{1\text{c}}^-| \hat{p}_y |\Gamma_{2\text{v}}^+\rangle.
\label{eq:Py4}
\end{align}

Under the basis functions of $\{\Gamma_{1\text{c}}^-,\Gamma_{2\text{c}}^+,\Gamma_{2\text{v}}^+,\Gamma_{1\text{v}}^-\}$, one could construct a $4\times4$  Hamiltonian similar to Eq.~(\ref{eq:H_3by3}), including all important $k_y\hat{p}_y$ interactions affecting the valence band. However, just from the  Hamiltonian matrix element analysis in Eqs.(\ref{eq:Py2})-(\ref{eq:Py4}), the application of L\"owdin partitioning already gives the unusual valence band dispersion in the $k_y$ direction approximated by 
\begin{align}
E_h(k_y) &= \frac{\hbar^2}{2m_0}k_y^2-\frac{P_{y4}^2}{E_{3}}k_y^2-\frac{P_{y2}^2 k_y^2}{E_{1v}(k_y)},
\label{eq:E_h_ky} 
\end{align}
where $E_3$ is the energy difference from $\Gamma_{1\text{c}}^-$ to $\Gamma_{2\text{v}}^+$. Note that, in the denominator of the last term, we use the complete dispersion of the $\Gamma_{1\text{v}}^-$ band [Eq.~(\ref{eq:E_1v})] which is negative and depends on $k_y$, rather than -$E_2$, the fixed energy difference between $\Gamma_{1\text{v}}^-$ and $\Gamma_{2\text{v}}^+$.

Eq.~(\ref{eq:E_h_ky}) guarantees a hole-like dispersion if the second term (repulsion from $\Gamma_{1\text{c}}^-$) overcomes the first (free electron) term. The ultraflat valence band along $k_y$ is thus due to the counteracting effects of $\Gamma_{1\text{v}}^-$ and $\Gamma_{1\text{c}}^-$ close to the zone center; phosphorene is indirect-gap ($k_0\neq 0$) when the matrix element magnitudes satisfy $\frac{P_{y2}^2}{E_{2}}+\frac{\hbar^2}{2m_0}>\frac{P_{y4}^2}{E_{3}}>\frac{\hbar^2}{2m_0}$. Although this condition is predicted by DFT, other indications of inaccuracies or variations in that numerical method (for example, gross underestimation of the bandgap or the details of relaxed structure)\cite{Tran_PRB14_2} suggest that we cannot exclude the possibility that repulsion from $\Gamma_{1\text{c}}^-$ is so strong that $\frac{P_{y4}^2}{E_{3}}>\frac{P_{y2}^2}{E_{2}}+\frac{\hbar^2}{2m_0}$, resulting in a direct gap. The true nature of the valence band must therefore be revealed by experiment.

\subsection{Optical selection rules}
The interaction between matter and radiation $\frac{e}{m}\mathbf{ A\cdot \hat p}$ has the same symmetry as $H_{\mathbf{k}\cdot\mathbf{\hat{p}}}$ (the electric field of light transforms like an ordinary polar vector). Therefore, we can use the perturbative coupling superimposed as arrows on Fig.~\ref{fig:bands} to reveal the selection rules of optical transitions between conduction and valence bands in phosphorene. 

The dominant transition paths are represented by the two red double arrows. When the electric field polarization is parallel to the $x$-direction, photons with $\hbar\omega\geq E_g$ can cause excitation across the bandgap from $\Gamma_{2\text{v}}^+$ to $\Gamma_{4\text{c}}^-$. On the other hand, photons with orthogonal polarization parallel to the $y$-direction can excite lower valence band $\Gamma_{1\text{v}}^-$ electrons to the upper conduction band $\Gamma_{2\text{c}}^+$, which requires a frequency deep into the ultraviolet regime ($E_g+E_1+E_2\sim5$ eV photon energy). In addition to the bands shown in Fig.~(\ref{fig:bands}) that are relevant to the bandedge dispersion, we note that there are also a $\Gamma_{4\text{v}}^+$ band between $\Gamma_{2\text{v}}^+$ and $\Gamma_{1\text{v}}^-$, as well as a $\Gamma_{3\text{c}}^-$ band in between $\Gamma_{2\text{c}}^+$ and $\Gamma_{3\text{c}}^+$. Optical transition with $y$-polarization is also allowed between these two states ($\Gamma_{4\text{v}}^+\otimes \Gamma_{3\text{c}^-}=\Gamma_{2}^-$), which are seperated by $\sim5$~eV but extend along the $\Gamma-X$ axis and reach a minimum energy separation at the location of a valence band satellite valley. This could explain the $\sim3.7$~eV onset photon energy in the calculated absorption spectrum of $y$-polarized light.\cite{Tran_PRB14}

The sensitivity of the absorption spectrum to electric field orientation, confirmed by DFT calculation,\cite{Qiao_NatureComm2014} may make phosphorene an ideal photoconductive polarimeter in near-infrared frequencies. 

In anticipation of the next section where we consider spin properties, it is appropriate to point out in the present discussion of radiative selection rules that optical orientation in phosphorene is hopelessly inefficient: Unlike in bulk semiconductors with cubic symmetry, where degeneracy of the $p$-like valence band is preserved in the spin-independent Hamiltonian allowing strong mixing of spin and orbital angular momentum, here with inversion-symmetric but orthorhombic phosphorene we have only nondegenerate and energetically well-separated bands. Spin-orbit interaction can thus only have a small perturbative mixing of spin states. Illumination with linear polarization almost entirely preserves spin during absorptive transitions, so that excitation of the spin-unpolarized valence band will result only in population of a nearly spin-unpolarized conduction band.

\section{Spin-dependent properties}
\label{sec:IV}

The spin-orbit interaction (SOI) in semiconductors composed of light atoms is expected to be weak, resulting in relatively pure spin states and long spin lifetime in the intrinsic regime. By this reasoning, SOI in phosphorus allotropes should resemble that of its  neighbor on the periodic table, silicon, an indirect-gap group-IV bulk semiconductor that has a long spin lifetime for electrons in the conduction band\cite{Huang_PRL2007,Pengke_PRL11,Cheng_PRL10} and a small split-off energy (44~meV) in the valence band. The latter feature of the bandstructure is a result of the SOI perturbing an otherwise three-fold degenerate p-like valence band extremum. Phosphorene, on the other hand, has only non-degenerate and energy-isolated bands (other than the few accidental crossings). When SOI is included in the calculation, we therefore expect only that some of the bands will be unnoticeably shifted, as confirmed by DFT calculation. \cite{Qiao_NatureComm2014}

Although SOI has a negligible effect on the band structure (Hamiltonian eigenvalues) in phosphorene, it is still worthwhile to examine the effect it has on spin-dependent eigenstates, especially because the in-plane anisotropy is expected to extend to spin-related phenomena such as relaxation mechanisms. In this section, we will now derive the \emph{spin-dependent} Hamiltonian including $H_{\text{SO}}$ and $H_{\text{SO},{\mathbf{k}}}$, again using the method of invariants, and calculate the spin-dependent eigenstates of electrons and holes that capture the underlying symmetries.  

Before starting our discussion, we compare some other spin-related properties between phosphorene and silicon, the latter of which is believed to be a promising material candidate for spintronic devices.\cite{Jansen_NatureMat2012} Besides the low atomic number already mentioned, there are two important factors leading to the rather long spin lifetime in Si. One is the centrosymmetric property of the diamond lattice that (along with Kramers' time-reversal symmetry) preserves spin degeneracy, and precludes the Dyakonov-Perel spin relaxation mechanism where momentum scattering causes a fluctuating effective magnetic field driving spin flips.\cite{Dyakonov_SPSS1972} The other is the absence of nuclear spin in the most abundant isotope $^{28}$Si, so that there is little hyperfine interaction affecting the electron spin states. 

Whereas the high abundance of $^{31}$P isotope (with half-integer nuclear spin) makes hyperfine interaction in phosphorene worth considering, it is outside the scope of this paper. However, phosphorene is indeed centrosymmetric (see Table~\ref{tab:CharacterTable}), and so like Si,  Dyakonov-Perel spin relaxation is absent. The dominant spin relaxation mechanism in both materials is then the Elliott and Yafet processes.

Generally speaking, perturbative spin-orbit coupling between different bands causes a mixing of pure Pauli spinors so that the two degenerate spin states within the same $i$th band can be written as 
\begin{align}
|i\!\Uparrow\rangle&\!=\!
C_i\left(|i\!\uparrow\rangle+\sum_{j\neq i} a_j |j\!\uparrow\rangle+b_j |j\!\downarrow\rangle\right),{\text{ and}}
\label{eq:s_SO_u_g} 
\\
|i\!\Downarrow\rangle&\!=\!
C_i\left(|i\!\downarrow\rangle+\sum_{j\neq i} a_j^* |j\!\downarrow\rangle-b_j^* |j\!\uparrow\rangle\right),
\label{eq:s_SO_d_g} 
\end{align}
where $C_i$ is a normalization factor. The dominant orbital component in both $|i\!\Uparrow\rangle$ and $|i\!\Downarrow\rangle$ is $|i\rangle$, whose amplitude is very close to unity. However, the coefficients $|b_j|\leqq|a_j|\ll1$ depending on their perturbation origins. In our following discussion, we ignore the prefactor $C_i\approx 1$, and for clarity, write the spin-dependent eigenstates in the form of 2-row matrices such as
\begin{align}
|i\!\Uparrow\rangle&\!=\!
\left(\!\!
\begin{array}{ccccc}
a_1&   a_2   &...& 1 &  ... \\
b_1  & b_2  & ...&0 &...
\end{array}
\!\!\right), {\text{ and}}
\label{eq:s_SO_u_m} 
\\
|i\!\Downarrow\rangle&\!=\!
\left(\!\!
\begin{array}{ccccc}
-b_1^* &   -b_2^* &...  & 0 &... \\
a_1^*  & a_2^* &... & 1& ...
\end{array}
\!\!\right),
\label{eq:s_SO_d_m}
\end{align}
where quantities in the top (bottom) row indicate coefficients of $|\!\!\uparrow\rangle$ ($|\!\!\downarrow\rangle$) states, and each column corresponds to a given band.

Spin flips are induced by an interaction $\langle\Uparrow\!\!|\Xi|\!\!\Downarrow\rangle$, where $\Xi$ is a scattering potential. The spin-independent part of this potential, normally leading to momentum scattering, will then couple the same spin components of $|\!\!\Uparrow\rangle$ and $|\!\!\Downarrow\rangle$, resulting in the Elliot mechanism\cite{Elliott_PR1954}, whereas the spin-dependent part of the scatterer will couple opposite spin components and give rise to the Yafet term\cite{Yafet_SSP1963}. These two quantities coherently interfere to yield the total spin relaxation rate $\tau_s^{-1}$, proportional to the square of the matrix element according to Fermi's golden rule. A major goal of the following subsections is to examine the symmetry of these processes and understand the spin-relaxation anisotropy in terms of the spin-dependent eigenstate components.

\subsection{Hamiltonian and spin-dependent eigenstates}

The incorporation of spin-orbit interactions [Eqs.~(\ref{eq:H_SO}) and (\ref{eq:H_SOk})] within the framework of the method of invariants is as straightforward as our treatment of $H_{\mathbf{k}\cdot\mathbf{\hat p}}$. We start with a discussion of the valence band Hamiltonian matrix of the $k$-independent SOI ($H_{\text{SO}}$). From Table~\ref{tab:table_of_inv}, one finds that the three invariant components of $H_{\text{SO}}$ belong to the IRs of $\Gamma_2^+$, $\Gamma_3^+$ and $\Gamma_4^+$, which couples the highest valence band $\Gamma_{2\text{v}}^+$ to remote bands belong to $\Gamma_{1}^+$, $\Gamma_{4}^+$ and $\Gamma_{3}^+$ , respectively. Therefore, in a basis   \{$\Gamma_{3}^+$, $\Gamma_{4}^+$, $\Gamma_{1}^+$, $\Gamma_{2\text{v}}^+$\}, the matrix form of $H_{\text{SO}}$ can be written
\begin{equation}
H_{\text{SO}}^{\text{hole}}=
\left(
\begin{array}{ccc|c}
    &  &  &   i\delta_{32}\sigma_z            \\
    &    &  & i\delta_{42}\sigma_x    \\ 
     &   &    & i\delta_{12}\sigma_y  \\\hline 
       -i\delta_{32}\sigma_z   &    -i\delta_{42}\sigma_x &-i\delta_{12}\sigma_y & 
\end{array}
\right),
\label{eq:H_SO_h}
\end{equation}
where
\begin{equation}
 \delta_{l2} = \sum_{\Gamma_{l}^+}\frac{i\hbar}{4m_0^2c^2}\langle \Gamma_{l}^+| \frac{\partial V}{\partial x_m } \hat{p}_n-\frac{\partial V}{\partial x_n} \hat{p}_m |\Gamma_{2\text{v}}^+\rangle,
\label{eq:delta_l2}
\end{equation}
which are real numbers taking into account all remote bands belonging to the IR $\Gamma_{l}^+$. Here, \{$lmn$\} is a cyclic permutation of \{1,2,3\} and \{$x_1$,$x_2$,$x_3$\} correspond to \{$x$, $y$, $z$\}. By expanding the Pauli matrices, $H_{\text{SO}}^{\text{hole}}$ in Eq.~(\ref{eq:H_SO_h}) is an $8\times8$ matrix, where each basis function is the direct product of a spin-independent wave function and a spinor ($|\!\!\uparrow\rangle$ or $|\!\!\downarrow\rangle$, eigenstates of $\sigma_z$). Note that the lack of matrix elements other than those involving $\Gamma_{2\text{v}}^+$ in Eq.~(\ref{eq:H_SO_h}) does not imply that they are zero; rather, we focus here only on the matrix elements that are important in determining the valence band eigenvectors to lowest order.   

$H_{\text{SO},{\mathbf{k}}}$ in Eq.~(\ref{eq:H_SOk}) gives a small correction to $H_{\text{SO}}$ that is usually negligible, since $|\hbar\mathbf{k}|$ is comparable with $|\mathbf{p}|$ only when $\mathbf{k}$ reaches the zone edge. In phosphorene, however, due to the unusually flat valence band, $k_y$ of hole states can be relatively large ($\gtrsim10\%$ of $\pi/a_y$), and the resulting $H_{\text{SO},{\mathbf{k}}}$ cannot be ignored. The components of $H_{\text{SO},{\mathbf{k}}}$ transform like polar vectors as shown in Table~\ref{tab:table_of_inv} and for simplicity we only keep the more important $k_y$-related terms. Applying the same procedure as above, we obtain the matrix form of $H_{\text{SO},{\mathbf{k}}}$ for hole states as
\begin{equation}
H_{\text{SO},\mathbf{k}}^{\text{hole}}=
\left(
\begin{array}{c|cc}
   &    i\alpha_{42}k_y\sigma_z     &  -i \alpha_{32}k_y\sigma_x  \\\hline
-i\alpha_{42}k_y\sigma_z       &    &     \\        
i \alpha_{32}k_y\sigma_x      &    &   
\end{array}
\right),
\label{eq:H_SOk_h}
\end{equation}
which is a $6\times6$ matrix in the basis \{$\Gamma_{2\text{v}}^+$, $\Gamma_{4}^-$, $\Gamma_{3}^-$\}$\otimes$\{$\uparrow$, $\downarrow$\}. Again, we are only interested in the matrix elements that couple to $\Gamma_{2\text{v}}^+$ in lowest order. The $\alpha$ parameters in Eq. (\ref{eq:H_SOk_h}) are
\begin{align}
 \alpha_{32} &= \sum_{\Gamma_{3}^-}\frac{i\hbar^2}{4m_0^2c^2}\langle \Gamma_{3}^-| \frac{\partial V}{\partial z } |\Gamma_{2\text{v}}^+\rangle, {\text{ and}}\label{eq:alpha_32} \\
 \alpha_{42} &= \sum_{\Gamma_{4}^-}\frac{i\hbar^2}{4m_0^2c^2}\langle \Gamma_{4}^-| \frac{\partial V}{\partial x } |\Gamma_{2\text{v}}^+\rangle. \label{eq:alpha_42} 
\end{align}

The essential spin-orbit interaction of hole states is now captured by $H_{\text{SO}}^{\text{hole}}\oplus H_{\text{SO},\mathbf{k}}^{\text{hole}}$, reducible to a $12\times12$ matrix because of the redundant $\Gamma_{2\text{v}}^+$. In combination with the spin-independent $H_0$ and $H_{\mathbf{k}\cdot\mathbf{\hat{p}}}$ operators discussed in the previous section, one can diagonalize the total Hamiltonian and obtain the spin-dependent eigenstates for holes. 

Several simplifying approximations can be made. Because it appears in the same matrix element and is much smaller than the $P_{x1}k_x$ terms in Eq.~(\ref{eq:H_3by3}), $\alpha_{42}k_y\sigma_z$ in Eq.~(\ref{eq:H_SOk_h}) can reasonably be ignored. In addition, the $P_yk_y$ terms (coupling between $\Gamma_{2\text{v}}^+$ and $\Gamma_1^-$ bands) can be neglected for two reasons: (\emph{i.}) the amplitudes of $P_{y2}$ and $P_{y4}$ are small; and (\emph{ii.}) the effect from $\Gamma_1^-$ bands below and above $\Gamma_{2\text{v}}^+$ counteract each other near the $\Gamma$-point (reflected by the flat band there).

To avoid lengthy summations and energy denominators, we define the following quantities related to the $\delta$, $\alpha$  and $P$ parameters:
\begin{align}
 \Delta_{l2} &= \sum_{\Gamma_{l}^+}\frac{i\hbar}{4m_0^2c^2}\frac{\langle \Gamma_{l}^+| \frac{\partial V}{\partial x_m } \hat{p}_n-\frac{\partial V}{\partial x_n} \hat{p}_m |\Gamma_{2\text{v}}^+\rangle}{E_{\Gamma_{l}^+}-E_{\Gamma_{2\text{v}}^+}},
\label{eq:Delta_l2} \\
& A_{32}k_y = \sum_{\Gamma_{3}^-}\frac{i\hbar^2}{4m_0^2c^2}\frac{\langle \Gamma_{3}^-| \frac{\partial V}{\partial z } |\Gamma_{2\text{v}}^+\rangle k_y}{E_{\Gamma_{3}^-}-E_{\Gamma_{2\text{v}}^+}}, {\text{ and}}\label{eq:A_32} \\
&\Pi_{x1}k_x = \frac{P_{x1}k_x}{E_g}. \label{eq:Pi_x1}
\end{align}
Because of their origins in the $H_{\mathbf{k}\cdot\mathbf{\hat p}}\gg H_{\text{SO}} \gg H_{\text{SO},{\mathbf{k}}}$ perturbation terms, the hierarchy between these unitless parameters is $A_{32}k_y\ll\Delta_{l2}\ll\Pi_{x1}k_x\ll 1$.

Each spin-dependent eigenvector includes 12 coefficients corresponding to the 12 basis functions which span the subspace of six IRs \{$\Gamma_{3}^+$, $\Gamma_{4}^+$, $\Gamma_{1}^+$, $\Gamma_{2\text{v}}^+$, $\Gamma_{4\text{c}}^-$, $\Gamma_{3}^-$\} and the two spinors. We write the spin-dependent eigenvectors in the form of Eqs.~(\ref{eq:s_SO_u_m}) and (\ref{eq:s_SO_d_m}) giving $2\times6$ matrices: 
\begin{align}
|\text{h}\!\Uparrow_\perp\rangle&\!=\!
\left(
\begin{array}{cccccc}
i\Delta_{32}&    \!\!0   &  0   & \,\,\,1 & \,\,\Pi_{x1}k_x & \!\!0 \\
     0  &  \!\!i\Delta_{42}     & \Delta_{12} &  \,\, \,0  &\,\, 0& \!\!\!\!-iA_{32} k_y
\end{array}
\!\!\right), {\text{ and}}
\label{eq:s_SO_h_u} 
\\
|\text{h}\!\Downarrow_\perp\rangle&\!=\!
\left(\!\!
\begin{array}{cccccc}
     0  & \!\! \!\!i\Delta_{42}     &  -\Delta_{12} &   0  & 0& \!\!\!\!-iA_{32} k_y \\
-i\Delta_{32}&    \!\!\!\!0   &  \!\!0   & 1 & \Pi_{x1}k_x & \!\!\!\!0 
 \end{array}
\!\!\right).
\label{eq:s_SO_h_d}
\end{align}
Here `h' stands for `hole' and the subscript `$\perp$' indicates that the spin orientation $z$ is out of plane. Notice that, before the trivial normalization, the dominant coefficients have the value of 1,  corresponding to the $|\Gamma_{2\text{v}}^+\!\!\uparrow\rangle$ ($|\Gamma_{2\text{v}}^+\!\!\downarrow\rangle$) basis function in $|\text{h}\!\Uparrow_\perp\rangle$ ($|\text{h}\!\Downarrow_\perp\rangle$).

Eqs.~(\ref{eq:s_SO_h_u}) and (\ref{eq:s_SO_h_d}) explicitly indicate the spin-purity of the eigenstates. One can evaluate the total square amplitude of the minority-spin components (the so-called `spin mixing' coefficient), which is approximately $\Delta_{42}^2+\Delta_{12}^2$ (the $A_{42}^2k_y^2$ term is a small correction). In Eq.~(\ref{eq:Delta_l2}), the energy denominators are several eV or more, while the remainder are within the same order of the $\delta_{l2}$ parameters in Eq.~(\ref{eq:delta_l2}). It is known that the dominant contribution to spin-orbit coupling is from the part of wavefunctions orthogonal to the core states, which are populated in the vicinity of the nucleus where the atomic potential changes drastically.\cite{Liu_PR62,Weisz_PR66} As in Si, these core electrons are $2p$ states, so we can estimate using the same parameters: $\delta_{l2}\sim$meV.\cite{Pengke_PRL11} This value then gives $\Delta_{l2}\sim10^{-3}$, so the spin mixing coefficient for holes in phosphorene should be similar to that of conduction band electrons in Si, which is approximately $10^{-5}-10^{-6}$.

Eqs.~(\ref{eq:s_SO_h_u}) and (\ref{eq:s_SO_h_d}) are fundamental in helping us understand the spin-flip process, in which a transition occurs between $|\text{h}\!\Uparrow_\perp\rangle$ and $|\text{h}\!\Downarrow_\perp\rangle$ during momentum scattering. As an example, we analyze the symmetry of the Elliott-Yafet (EY) spin-flipping process due to scattering by small-$k$ long-wavelength acoustic phonons (in-phase quasi-uniform vibration of atoms). This is the dominant spin relaxation mechanism that limits the intrinsic spin-lifetime in such a centrosymmetric system at finite temperature. 

The Elliott term\cite{Elliott_PR1954} proportional to $\nabla V$ has the same symmetry of a polar vector or the $p$ orbitals of  $\Gamma_2^-$ in Fig.~\ref{fig:py} , $\Gamma_3^-$ in Fig.~\ref{fig:pxpzs}(a) and $\Gamma_4^-$ in Fig.~\ref{fig:pxpzs}(b). It couples the same spin components between $|\text{h}\!\Uparrow_\perp\rangle$ and $|\text{h}\!\Downarrow_\perp\rangle$. Eqs.~(\ref{eq:s_SO_h_u}) and (\ref{eq:s_SO_h_d}) show that this happens between $\Gamma_{2\text{v}}^+$ and $\Gamma_{3}^-$ (with the coefficients $1$ and $-A_{32} k_y$), as well as between $\Gamma_{4\text{c}}^-$ and $\Gamma_{1}^+$ (with the coefficients $\Pi_{x1} k_x$ and $\Delta_{12}$), and the responsible phonon mode is the $\Gamma_4^-$-related $\frac{\partial V}{\partial z}$ Elliott term (out-of-plane motion of atoms, or flexural phonons).

On the other hand, the Yafet operator that couples opposite spin components\cite{Yafet_SSP1963} between $|\text{h}\!\Uparrow_\perp\rangle$ and $|\text{h}\!\Downarrow_\perp\rangle$ is proportional to $\nabla H_{\text{SO}}$, which is the gradient of an axial vector  with symmetry
\begin{align}
 &(\Gamma_4^-\oplus\Gamma_3^-\oplus\Gamma_2^-)\otimes(\Gamma_4^+\oplus\Gamma_3^+\oplus\Gamma_2^+)\nonumber\\
 =&3\Gamma_1^-\oplus2\Gamma_2^-\oplus2\Gamma_3^-\oplus 2\Gamma_4^-.
 \label{eq:Yafet}
\end{align}
Examining Eqs.~(\ref{eq:s_SO_h_u}) and (\ref{eq:s_SO_h_d}), one sees that the dominant coupling is between $\Gamma_{2\text{v}}^+$ and $\Gamma_{4\text{c}}^-$ by one of the two $\Gamma_3^-$ IRs in Eq.~(\ref{eq:Yafet}) that is related to $\frac{\partial}{\partial z}\left(\frac{\partial V}{\partial z}p_x-\frac{\partial V}{\partial x}p_z\right)\sigma_y$. This term also corresponds to flexural phonons [the other term belonging to $\Gamma_3^-$ IR is $\frac{\partial}{\partial y}\left(\frac{\partial V}{\partial x}p_y-\frac{\partial V}{\partial y}p_x\right)\sigma_z$, which does not flip the spin].

The same formalism can be applied to study the spin-dependent conduction band electrons. Here, we list the expressions for Hamiltonian matrices and spin-dependent eigenstates of the lowest eigenvalue (conduction band minimum). In a basis \{$\Gamma_{3}^-$, $\Gamma_{2}^-$, $\Gamma_{1}^-$, $\Gamma_{4\text{c}}^-$\}, $H_{\text{SO}}$ for $\Gamma_{4\text{c}}^-$ electrons is
\begin{equation}
H_{\text{SO}}^{\text{electron}}=
\left(
\begin{array}{ccc|c}
    &   &   &    i\delta_{34}\sigma_y             \\
    &   &  &  i\delta_{24}\sigma_x       \\ 
    &     &   &    i\delta_{14}\sigma_z   \\\hline
-i \delta_{34}\sigma_y &      - i\delta_{24}\sigma_x & -i\delta_{14}\sigma_z &  
\end{array}
\right),
\label{eq:H_SO_e}
\end{equation}
where 
\begin{equation}
 \delta_{l4} = \sum_{\Gamma_{l}^-}\frac{\hbar^2}{4m_0^2c^2}\langle \Gamma_{l}^-| \frac{\partial V}{\partial x_m } \hat{p}_n-\frac{\partial V}{\partial x_n} \hat{p}_m |\Gamma_{4\text{c}}^-\rangle.
\label{eq:delta_l4}
\end{equation}
As in Eq.~(\ref{eq:H_SOk_h}) for holes, the $H_{\text{SO},{\mathbf{k}}}$ matrix for electrons in the $k_y$ direction, using the basis \{$\Gamma_{4\text{c}}^-$, $\Gamma_{2}^+$, $\Gamma_{1}^+$\}, reads
\begin{equation}
H_{\text{SO},\mathbf{k}}^{\text{electron}}=
\left(
\begin{array}{c|cc}
   &    i\alpha_{24}k_y\sigma_z     &  - i\alpha_{14}k_y\sigma_x  \\\hline
-i\alpha_{24}k_y\sigma_z       &    &     \\        
i \alpha_{14}k_y\sigma_x      &    &   
\end{array}
\right),
\label{eq:H_SOk_e}
\end{equation}
where the $\alpha$ parameters are
\begin{align}
 \alpha_{24} = \sum_{\Gamma_{2}^+}\frac{i\hbar^2}{4m_0^2c^2}\langle \Gamma_{2}^+| \frac{\partial V}{\partial z } |\Gamma_{4\text{c}}^-\rangle, {\text{ and}}\label{eq:alpha_24} \\
 \alpha_{14} = \sum_{\Gamma_{1}^+}\frac{i\hbar^2}{4m_0^2c^2}\langle \Gamma_{1}^+| \frac{\partial V}{\partial x } |\Gamma_{4\text{c}}^-\rangle. \label{eq:alpha_14} 
\end{align}
As in Eqs.~(\ref{eq:Delta_l2}) and (\ref{eq:A_32}), we define 
\begin{align}
 \Delta_{l4} &= \sum_{\Gamma_{l}^-}\frac{i\hbar}{4m_0^2c^2}\frac{\langle \Gamma_{l}^+| \frac{\partial V}{\partial x_m } \hat{p}_n-\frac{\partial V}{\partial x_n} \hat{p}_m |\Gamma_{4\text{c}}^-\rangle}{E_{\Gamma_{l}^-}-E_{\Gamma_{4\text{c}}^-}},{\text{ and}}
\label{eq:Delta_l4} \\
& A_{14}k_y = \sum_{\Gamma_{1}^+}\frac{i\hbar^2}{4m_0^2c^2}\frac{\langle \Gamma_{1}^+| \frac{\partial V}{\partial z } |\Gamma_{4\text{c}}^-\rangle k_y}{E_{\Gamma_{1}^+}-E_{\Gamma_{4\text{c}}^-}}. \label{eq:A_14} 
\end{align}
The resulting spin-dependent electron eigenstates in the basis \{$\Gamma_{3}^-$, $\Gamma_{2}^-$, $\Gamma_{1}^-$, $\Gamma_{4\text{c}}^-$, $\Gamma_{2\text{v}}^+$, $\Gamma_{1}^+$\} are
\begin{align}
|\text{e}\!\Uparrow_\perp\rangle&\!=\!
\left(
\begin{array}{cccccc}
0   &  0   &\,\,i\Delta_{14}& \,1 & \,\,\Pi_{x1}k_x & \!\!0 \\
\Delta_{34}     & i\Delta_{24} &      \,\,0  &   \,0  &\,\, 0& \!\!\!-iA_{14} k_y
\end{array}
\!\!\right), {\text{ and}}
\label{eq:s_SO_e_u} 
\\
|\text{e}\!\Downarrow_\perp\rangle&\!=\!
\left(\!\!
\begin{array}{cccccc}
-\Delta_{34}     &  i\Delta_{24} & \!\!0 &   0  & 0& \!\!\!-iA_{14} k_y \\
0   &  0   & \!\!-i\Delta_{14}&   1 & \Pi_{x1}k_x & \!\!\!0 
 \end{array}
\!\!\right).
\label{eq:s_SO_e_d}
\end{align}

\subsection{Anisotropy of spin-dependent properties}

We naturally expect that the orthorhombic inequivalence of the three orthogonal armchair, zigzag, and out-of-plane axes will induce anisotropy of the spin-dependent properties of electrons and holes. Already from our analysis of the band dispersion, we can see an obvious anisotropy in the spin-diffusion length $\lambda_s = \sqrt{\tau_s D}$, where the carrier diffusion coefficient $D$ is related to the anisotropic effective mass. However, there is an additional contribution: the spin lifetime $\tau_s$ also has an anisotropy, related in this case not to the wavevector direction, but rather to the spin orientation.

In the previous subsection, we have derived the spin-dependent Hamiltonian and eigenstates with the spin orientation in the out-of-plane $z$-direction. However, in most spin-injection experiments, the spin orientation is fixed by the magnetization of ferromagnetic thin-film contacts with an in-plane easy axis. It is therefore more relevant to study the spin-dependent eigenstates and related properties with the spin quantization axis $z$ in-plane. Of course, the spatial symmetry of the system is invariant regardless of the coordinate system we choose, and so are the properties of IRs and the relations between them (Table \ref{tab:directproduct}); only the coordinate labels change. 

First, we calculate the spin-dependent properties of hole states under a new coordinate system where $z$ is chosen to be along the armchair, $x$ along the zigzag and $y$ along the out-of-plane direction. We apply the cyclic permutation $xyz\rightarrow zxy$ to the coordinates in Fig.~\ref{fig:lattice}(a), and the corresponding invariant components in Table~\ref{tab:table_of_inv} (note that one could also keep the coordinate system unchanged but alternatively derive spin-dependnet eigenstates from linear combination of $|\text{h}\!\Uparrow_\perp\rangle$ and $|\text{h}\!\Downarrow_\perp\rangle$, and $|\text{e}\!\Uparrow_\perp\rangle$ and $|\text{e}\!\Downarrow_\perp\rangle$, according to the chosen spin orientation). The spin-dependent Hamiltonian can be derived straightforwardly, so we do not repeat the previous procedures but rather give the result of the spin-dependent eigenstates.

Like Eqs.~(\ref{eq:s_SO_h_u}) and (\ref{eq:s_SO_h_d}), the spin-dependent eigenvectors for valence-band holes under the new coordinate system also includes 12 coefficients corresponding to the 12 basis functions which expand the subspace of six IRs \{$\Gamma_{3}^+$, $\Gamma_{4}^+$, $\Gamma_{1}^+$, $\Gamma_{2\text{v}}^+$, $\Gamma_{4\text{c}}^-$, $\Gamma_{3}^-$\} and the two spinors. In the form of Eqs.~(\ref{eq:s_SO_u_m}) and (\ref{eq:s_SO_d_m}), they read
\begin{align}
|\text{h}\!\Uparrow_{\text{ac}}\rangle&\!=\!
\left(
\begin{array}{cccccc}
\!\!0   \!&i\Delta_{42}&    0   &   \,\,1 &\, \Pi_{z1}k_z & \!-iA_{32} k_x \\
\!\Delta_{32} &     \!\!0  & i \Delta_{12}     &  \,\,0    &\, A_{42} k_x & \!0 
\end{array}
\!\!\right), {\text{ and}}
\label{eq:s_SO_h_u_ac} 
\\
|\text{h}\!\Downarrow_{\text{ac}}\rangle&\!=\!
\left(\!\!
\begin{array}{cccccc}
-\Delta_{32} &     \!\!\!0  &  \!\! i\Delta_{12}     &     0  & -A_{42} k_x& \! 0   \\
\!\!0   &\!\!\!-i\Delta_{42}&    \!\!0   &   1 & \Pi_{z1}k_z &  \!iA_{32} k_x  
 \end{array}
\!\!\right). 
\label{eq:s_SO_h_d_ac}
\end{align}
Here, the subscript `${\text{ac}}$' stands for `armchair' and indicates the spin orientation. $\Pi_{z1}$ has the same value as $\Pi_{x1}$ defined by Eq.~(\ref{eq:Pi_x1}) since we have only changed the labeling from $x$ to $z$. Also, similar to the definition of $A_{32}$ in Eq.~(\ref{eq:A_32}), we have
\begin{equation}
A_{42}k_y = \sum_{\Gamma_{4}^-}\frac{i\hbar^2}{4m_0^2c^2}\frac{\langle \Gamma_{4}^-| \frac{\partial V}{\partial z } |\Gamma_{2\text{v}}^+\rangle k_y}{E_{\Gamma_{4}^-}-E_{\Gamma_{2\text{v}}^+}}. \label{eq:A_42}
\end{equation}

Eqs.~(\ref{eq:s_SO_h_u_ac}) and ({\ref{eq:s_SO_h_d_ac}) directly show that spin-mixing to lowest order is $\Delta_{32}^2+\Delta_{12}^2$, different from the previous case ($\Delta_{42}^2+\Delta_{12}^2$) when the spin orientation $z$ is out-of-plane. It will be shown that if $z$ is along the zigzag direction, the spin mixing is $\Delta_{32}^2+\Delta_{42}^2$ [see Eqs.~(\ref{eq:s_SO_h_u_zz}) and {\ref{eq:s_SO_h_d_zz})]. The anisotropy of the spin-mixing is generally reflected in the definition of Eq.~(\ref{eq:Delta_l2}) by the energy denominators, while the numerators scale similarly, due to their major origins related to the $2p$ core states, as previously discussed.\cite{Liu_PR62, Weisz_PR66} Numerical calculation shows that, for the $\Gamma_{2\text{v}}^+$ hole band, the closest $\Gamma_3^+$ is the fourth conduction band approximately 3.5~eV above, while the closest neighboring $\Gamma_4^+$ and $\Gamma_1^+$ bands are the second and third valence bands (not shown in Fig.~\ref{fig:bands}) that are approximately 1.8~eV below. We therefore estimate that the spin mixing with $z$ in-plane is approximately half of that with $z$ out-of-plane, and the Elliott-process-limited spin lifetime is consequently $\sim 4$ times longer.

A more intriguing feature of the anisotropy is that of the phonon polarization in spin-flip scattering. Again, we take the interaction with long wavelength acoustic phonons as an example. In contrast to the case when $z$ is out-of-plane, here the coupling of the dominant $\Gamma_{2\text{v}}^+$ band by the Elliott operator to $\Gamma_3^-$ vanishes, since the coefficient for $\Gamma_3^-\!\!\uparrow (\downarrow)$ in $|h\!\!\Downarrow_{ac}(\Uparrow_{ac})\rangle$ is zero. However, the coupling between $\Gamma_{2\text{v}}^+$ and $\Gamma_4^-$ survives, with the coefficient  $-A_{42} k_y$. More importantly, the interaction belongs to $\Gamma_3^-=\Gamma_{2\text{v}}^+\otimes\Gamma_{4}^-$, which still corresponds to $\frac{\partial V}{\partial z}$, but is now related to in-plane, instead of flexural, phonons. 

It is well-known that, due to their quadratic dispersion relation, the thermal population of flexural phonons in 2D materials diverges when the phonon wavevector $\mathbf{q}$ approaches zero. Such a singularity does not exist in the case of in-plane phonons with linear dispersion. The intrinsic spin lifetime in 2D materials is thus essentially limited by the interaction with flexural phonons while the influence from in-plane phonons is much less crucial.\cite{Song_PRL2013} However, here in phosphorene, for spins oriented in the armchair direction the Elliott spin-flip coupling of the dominant $\Gamma_{2\text{v}}^+$ component via flexural phonons is excluded by symmetry.

We have seen that scattering by acoustic phonons ultimately determines the upper bound of the spin lifetime in phosphorene. However, this intrinsic mechanism will be superseded by various extrinsic spin relaxation mechanisms in anything other than the most pure, well-isolated samples. Carrier spins in 2D materials are especially sensitive to extrinsic effects such as interactions with substrate and contacts, scattering with impurities or defects, and the influence of deformation such as nanoripples and strain. As long as these spin-flip processes are within the perturbative regime, the spin-dependent eigenvectors can be effectively used to determine whether a certain process is symmetry-allowed, as well as its orientation dependence. The strength of the interaction can be evaluated by the coefficients in these eigenvectors, calculated with the assistance of numerical schemes able to yield the coupling amplitude between basis functions.

Before closing this section, we derive expressions for the spin-dependent hole eigenvectors for the remaining orientation with $z$ along the zigzag direction by applying the permutation $xyz\rightarrow yzx$ to the table of invariants. Using a basis \{$\Gamma_{3}^+ $, $\Gamma_{4}^+$, $\Gamma_{1}^+$, $\Gamma_{2\text{v}}^+$, $\Gamma_{4\text{c}}^-$, $\Gamma_{3}^-$\}, the eigenvectors read
\begin{align}
|\text{h}\!\Uparrow_{\text{zz}}\rangle&\!=\!
\left(
\begin{array}{cccccc}
\!\!0   \!&0&    i\Delta_{12}   &   \,\,1 & \Pi_{y1}k_y & \!0 \\
\!i\Delta_{32} &  \Delta_{42}  &  0     &  \,\,0    & iA_{42} k_z & \!-A_{32} k_z 
\end{array}
\!\!\right), {\text{ and}}
\label{eq:s_SO_h_u_zz} 
\\
|\text{h}\!\Downarrow_{\text{zz}}\rangle&\!=\!
\left(\!\!
\begin{array}{cccccc}
i\Delta_{32} &     \!\!-\Delta_{42}  & \!\!\! 0     &     0  & iA_{42} k_z& \! A_{32} k_z   \\
0   & \!\!0&   \!\!\! -i\Delta_{12}   &   1 & \Pi_{y1}k_y &  \!0  
 \end{array}
\!\!\right).
\label{eq:s_SO_h_d_zz}
\end{align}
The subscript `${\text{zz}}$' stands for `zigzag'. Also, we give the spin-dependent eigenvectors of the conduction-band electron states for the spin orientation aligned with both in-plane axes, in a basis \{$\Gamma_{3}^-$, $\Gamma_{2}^-$, $\Gamma_{1}^-$, $\Gamma_{4\text{c}}^-$, $\Gamma_{2\text{v}}^+$, $\Gamma_{1}^+$\}. For $z$ along the armchair direction, the eigenvectors are

\begin{align}
|\text{e}\!\Uparrow_{ac}\rangle&\!=\!
\left(
\begin{array}{cccccc}
0   &  i\Delta_{24}   &0& 1 & \Pi_{z1}k_z & \!\!-iA_{14} k_x \\
i\Delta_{34}     & 0 & \Delta_{14}  &   0  & -A_{24} k_x& \!\!\!0
\end{array}
\!\!\right), {\text{ and}}
\label{eq:s_SO_e_u_ac} 
\\
|\text{e}\!\Downarrow_{ac}\rangle&\!=\!
\left(\!\!
\begin{array}{cccccc}
i\Delta_{34}     &  0 & \!\!-\Delta_{14} &   0  & A_{24} k_x& 0 \\
0   &  -i\Delta_{24}   & 0&   1 & \Pi_{x1}k_x & iA_{14} k_y 
 \end{array}
\!\!\right).
\label{eq:s_SO_e_d_ac}
\end{align}
For $z$ along the zigzag direction, they are
\begin{align}
|\text{e}\!\Uparrow_{zz}\rangle&\!=\!
\left(
\begin{array}{cccccc}
i\Delta_{34}   &  0   &0& \,1 & \,\,\Pi_{y1}k_y & 0 \\
0     & \Delta_{24} & i\Delta_{14}  &   \,0  &iA_{24} k_z & A_{14} k_z
\end{array}
\!\!\right), {\text{ and}}
\label{eq:s_SO_e_u_zz} 
\\
|\text{e}\!\Downarrow_{zz}\rangle&\!=\!
\left(\!\!
\begin{array}{cccccc}
0     &  \!\!\!-\Delta_{24} & i\Delta_{14} &   0  & iA_{24} k_z & \!\!-A_{14} k_z \\
-i\Delta_{34}   &  \!\!\!0   & 0 &   1 & \Pi_{y1}k_y & \!\!0 
 \end{array}
\!\!\right).
\label{eq:s_SO_e_d_zz}
\end{align}

\section{Summary}
\label{sec:V}

We have detailed the myriad ways in which phosphorene's structural in-plane asymmetry is manifest in the anisotropy of charge and spin properties of its electrons and holes, all of which are otherwise obscured by the many numerical methods previously applied to the problem. We have elucidated the origin of the band dispersion anisotropy near the gap edge, which is due to the specific directional preference of the coupling between conduction and valence bands and governs various transport and optical properties. By analyzing the symmetry of spin-orbit coupling, we have derived compact spin-dependent eigenstates of electrons and holes for all high-symmetry quantization axes. As an example of the utility of these eigenstates, we investigated the valence band anisotropy of the Elliott-Yafet spin-relaxation process. By incorporating the relevant invariant components into our model, such as those of  external applied fields, mechanical strain, and quantum confinement, this theory is highly extensible to the analysis of many other relevant circumstances of interest.

Importantly, our theory provides guidance for experimental efforts to empirically confirm and quantify these charge and spin phenomena. Valence band properties like the effective mass anisotropy and possible indirect bandgap are most directly compared to results from ARPES.\cite{Xia_NatureComm2014,Li_NatureNano2014} This ultra-high vacuum surface-sensitive method is especially convenient since electrical contact to phosphorene transport layers has not yet been realized, in part due to environmental sensitivity and subsequent degradation.\cite{Liu_ACSNano2014} Even without metallic electrodes, basic transport properties can be determined with e.g. microwave Hall mobility measurement.\cite{Portis_JAP1958} However, investigation into many other properties (such as thermal and excitonic transport, weak localization or antilocalization, etc) may require fabrication of true electronic devices. 

Optical non-contact techniques are also very useful in verifying the linear polarimetry in the visible spectrum, enabled by the dipole selection rules. However, with the expected inefficiency of optical orientation in this material, electrical techniques are required to measure anisotropy of the spin relaxation. Once single phosphorene layers can be stabilized for device processing and fabrication, four-terminal nonlocal geometry devices in the presence of oblique magnetic fields (to precess spins out-of-plane) will be especially applicable.\cite{Han_PRL10}

In closing, we are obliged to point out the presence of secondary features of the conduction and valence bands that cannot be captured simply by the zone-center symmetries. For example, satellite valleys may play a role in transport properties of phosphorene when a high electric field accelerates and heats mobile charge carriers. Interestingly, secondary conduction band valleys are predicted by DFT calculation along the $k_y$ direction, whereas the valence band valleys are along the $k_x$ direction, several hundred meV from the band edge.

\acknowledgments{We thank Dr. Yang Song for carefully reading the manuscript and providing valuable suggestions. We gratefully acknowledge support from the Office of Naval Research under contract N000141410317, the National Science Foundation under contract ECCS-1231855, and the Defense Threat Reduction Agency under contract HDTRA1-13-1-0013.}

%

\end{document}